\documentclass[fleqn,10pt]{wlscirep}

\usepackage{verbatim}
\usepackage{dsfont}
\usepackage{cleveref}
\usepackage{braket}
\usepackage{epstopdf}
\usepackage{cancel}

\newcommand{\tr}{\text{Tr}}

\newcommand{\av}[1]{\left\langle #1 \right\rangle}
\newcommand{\beq}{\begin{equation}}
\newcommand{\eeq}{\end{equation}}

\newcommand{\ip}[2]{\left\langle #1 | #2\right\rangle}

\begin{document}

\title{The quantum Zeno and anti-Zeno effects with non-selective measurements}

\author[1]{Mehwish Majeed}
\author[2,*]{Adam Zaman Chaudhry}
\affil[1,2]{School of Science \& Engineering, Lahore University of Management Sciences (LUMS), Opposite Sector U, D.H.A, Lahore 54792, Pakistan}

\affil[*]{adam.zaman@lums.edu.pk}

\begin{abstract}

In studies of the quantum Zeno and anti-Zeno effects, it is usual to consider rapid projective measurements with equal time intervals being performed on the system to check whether or not the system is in the initial state. These projective measurements are selective measurements in the sense that the measurement results are read out and only the case where all the measurement results correspond to the initial state is considered in the analysis of the effect of the measurements. In this paper, we extend such a treatment to consider the effect of repeated non-selective measurements - only the final measurement is required to correspond to the initial state, while we do not know the results of the intermediate measurements. We present a general formalism to derive the effective decay rate of the initial quantum state with such non-selective measurements. Importantly, we show that there is a difference between using non-selective measurements and the usual approach of considering only selective measurements only if we go beyond the weak system-environment coupling regime in models other than the usual population decay models. As such, we then apply our formalism to investigate the quantum Zeno and anti-Zeno effects for three exactly solvable system-environment models: a single two-level system undergoing dephasing, a single two-level system interacting with an environment of two-level systems, and a large spin undergoing dephasing. Our results show that the quantum Zeno and anti-Zeno effects in the presence of non-selective measurements can differ very significantly as compared to the repeated selective measurement scenario.

\end{abstract}

\maketitle

\section*{Introduction}
If a quantum system is subjected to repeated projective measurements, then the evolution of the quantum system slows down. This effect is known as the quantum Zeno effect (QZE) \cite{Sudarshan1977}. On the other hand, a more ubiquitous phenomenon under realistic conditions is the opposite effect - the acceleration of the quantum state evolution via the repeated measurements, known as the quantum anti-Zeno effect (QAZE) \cite{KurizkiNature2000, KoshinoPhysRep2005}. Both the QZE and the QAZE have attracted considerable attention \cite{ItanoPRA1994,FacchiPhysLettA2000,RaizenPRL2001,FacchiPRL2002,BaronePRL2004,ManiscalcoPRL2006,SegalPRA2007,FacchiJPA2008,
WangPRA2008,ManiscalcoPRL2008,ZhengPRL2008,AiPRA2010,FacchiJPA2010,BennettPRB2010,YamamotoPRA2010,ThilagamJMP2010, MilitelloPRA2011,XuPRA2011,ZhangJETP2011,CaoPhysLettA2012,RaimondPRA2012, SmerziPRL2012, WangPRL2013,McCuskerPRL2013,ThilagamJCP2013,ChaudhryPRA2014zeno,StannigelPRL2014, ZhuPRL2014, SchafferNatCommun2014,SignolesNaturePhysics2014, DebierrePRA2015, AlexanderPRA2015, QiuSciRep2015, FanSciRep2015, SlichterNJP2016,Chaudhryscirep2016,WuPRA2017,Chaudhryscirep2017a,Chaudhryscirep2017b,HanggiNJP2018,WuAnnals2018} and studies have been performed by considering a variety of experimental setups such trapped ions \cite{ItanoPRA1994}, cold atomic gas \cite{RaizenPRL2001}, nanomechanical oscillators \cite{BennettPRB2010}, and superconducting qubits \cite{SlichterNJP2016}. The general scenario - see, for example, Refs.~{\renewcommand{\citemid}{}\cite[]{KurizkiNature2000, KoshinoPhysRep2005,BennettPRB2010,BaronePRL2004,YamamotoPRA2010,RaizenPRL2001,ManiscalcoPRL2006,SegalPRA2007,ZhengPRL2008,
AiPRA2010,ThilagamJMP2010,ThilagamJCP2013}} - is to prepare initially an excited state of the system. This excited state then decays due to the system's interaction with the surrounding environment. The idea is to repeatedly check via repeated projective measurements whether or not the system is still in the excited state or not. Each projective measurement prepares the initial state, and any other measurement result is rejected. This scenario can be generalized to go beyond such population decay models in the sense that dephasing can also be taken into account \cite{ChaudhryPRA2014zeno}, and arbitrary system-environment models can be considered \cite{Chaudhryscirep2016}. 

In this paper, we go beyond such selective projective measurements usually considered in the analysis of the QZE and the QAZE. First, we consider `unsuccessful' measurements as well. In this case, we do read off the measurement results of the projective measurements, but we do not require the measurement results to correspond to the initial state for every measurement. Only the final measurement is required to do so. Second, rather than performing selective measurements, we can consider non-selective measurements where we do perform the measurements, but we do not read out the measurement results. Once again, only the final measurement is required to be a selective measurement corresponding to the initial state. A similar measurement strategy has been followed before to study the quantum Zeno and anti-Zeno effects for a harmonic oscillator coupled to a harmonic oscillator environment \cite{HanggiNJP2018}. For both of the above scenarios, the same final survival probability is obtained. In particular, we show that our expression for the final survival probability reduces to the expression obtained using the usual repeated selective projective measurement scheme if the system-environment coupling is evaluated using only first-order time-dependent perturbation theory and higher order terms are neglected. Our work is therefore a rare example of an investigation of the QZE and the QAZE beyond the weak system-environment coupling regime \cite{WuPRA2017,Chaudhryscirep2017a}. As a consequence, the usual perturbative techniques cannot be used and we use exactly solvable models to analyze the effect of the non-selective measurements. We consider three such models. First, we consider a single two-level system undergoing dephasing via its interaction with an environment consisting of harmonic oscillators. Second, we consider a single two-level system interacting with an environment consisting of many two-level systems. Third, we consider a large spin (or, equivalently, more than one two-level system) interacting with an environment of harmonic oscillators and undergoing dephasing. Using the expression for the final survival probability, we can define an effective decay rate in analogy with the usual studies of the quantum Zeno and anti-Zeno effects. The behavior of the effective decay rate allows us to investigate the effect of performing non-selective measurements instead of the usual selective measurement scheme. We show that the QZE and QAZE are considerably modified. In particular, the QZE and the QAZE effects now depend on the number of measurements performed. The effective decay rates are now reduced; moreover, the measurement rates corresponding to the crossover from the QZE regime to the QAZE regime and vice versa can also change.    

\section*{Results}
\subsection*{Background}
Before presenting our results, it is useful to recap the basic theory \cite{Chaudhryscirep2016}. The approach usually followed is that at initial time $t = 0$, the system quantum state $\rho_0$ is prepared. The system then interacts with the environment and evolves for time $\tau$ to the state $\rho_0(\tau)$. A projective measurement is then performed at time $\tau$ to check whether or not the system is still in the state $\rho_0$. Let this probability be $s_{00}$. We also note that since we are interested in the system evolution due to the system-environment interaction only, the evolution due to the free system Hamiltonian is removed just before performing the projective measurement by applying a suitable unitary operator on a very short timescale \cite{MatsuzakiPRB2010,ChaudhryPRA2014zeno,Chaudhryscirep2016,Chaudhryscirep2017a,Chaudhryscirep2017b}. The system state is then reset to $\rho_0$, and following another time interval $\tau$, another measurement is performed. The probability that the system is still in the initial state $\rho_0$ is $S(M\tau) = s_{00}^M$ if system-environment correlation effects are neglected. We can then define an effective decay rate $\Gamma(\tau)$ via $S(M\tau) = e^{-\Gamma(\tau)M\tau}$. In this case, $\Gamma(\tau)$ is then found to be $-\frac{1}{\tau}\ln s_{00} = -\frac{1}{\tau}\ln (1 - s_{01})$, where $s_{01}$ is the probability that the system, after a measurement, ends up in a state $\rho_1$ orthogonal to the initial state $\rho_0$. For weak system-environment coupling strength, we expect the transition probability $s_{01}$ to be small, leading to $\Gamma(\tau) \approx s_{01}/\tau$. The probability $s_{01}$ can then be calculated perturbatively to show that the effective decay rate depends on the overlap of the spectral density of the environment and an `effective' filter function that depends on the measurements performed, the measurement interval, and the system-environment model being considered \cite{Chaudhryscirep2016}. The effective decay rate $\Gamma(\tau)$ can then be plotted as a function of the measurement interval $\tau$. When $\Gamma(\tau)$ is an increasing function of $\tau$, we are in the Zeno regime, since in this case, decreasing the measurement interval decreases the effective decay rate. If the opposite is true, then we are in the anti-Zeno regime \cite{KurizkiNature2000, SegalPRA2007, ThilagamJMP2010, ChaudhryPRA2014zeno,Chaudhryscirep2016,Chaudhryscirep2017a,Chaudhryscirep2017b} . 

\subsection*{The formalism}
We now modify the scheme presented above to first take into account `unsuccessful' measurement results as well. We no longer demand that every measurement result corresponds to the initial state. Intermediate measurement results can correspond to state(s) other than the initial state - these measurements are what we refer to as unsuccessful measurements. We keep track of the result of every measurement, and only the final measurement result should correspond to the initial state. For simplicity, we consider here the case of a two-level system - higher dimensional systems can be treated in a similar manner as done later when we study the large spin pure dephasing model. The two-level system is initially prepared in the state $\rho_0$. We now perform repeated measurements on the system with time interval $\tau$ to check the state of the quantum system. Just after each measurement, the state of the system could be $\rho_0$, or it could be the state $\rho_1$, which is orthogonal to $\rho_0$, due to the system's interaction with the environment. As before, $s_{01}$ as the transition probability that the system ends up in state $\rho_1$ if it started in state $\rho_0$. In a similar manner, we can define $s_{10}$ $(s_{11})$ as the transition probability that the system ends up in state $\rho_0$ $(\rho_1)$ if it started in state $\rho_1$. We are interested in what happens after $M$ measurements; that is, what is the probability that the system is still in state $\rho_0$ after $M$ measurements? Calling this probability $S(M\tau)$, if we neglect any system-environment correlation effects, we can write
\begin{equation}
S(M\tau) = \sum_{i_1 i_2 \hdots i_{M - 1}} s_{0i_1}s_{i_1 i_2}s_{i_2 i_3} \hdots s_{i_{M - 2}i_{M - 1}} s_{i_{M - 1}0}. 
\label{finalprobability}
\end{equation}
This probability can be further evaluated using matrix multiplication (see the Methods section). The final result is
\begin{equation}
S(M\tau) = \frac{s_{01}(1 - s_{01} - s_{10})^M + s_{10}}{s_{01} + s_{10}}. $$
\label{result1}
\end{equation}
We emphasize that this result is independent of the details of the system-environment model - the only assumption is that the system-environment coupling is not so strong that system-environment correlation effects become very significant \cite{ChaudhryPRA2014zeno}. This expression can also be cast in a more illuminating form. Noting that 
$$ (1 - s_{01} - s_{10})^M = 1 + \sum_{k = 1}^M (-1)^M \binom{M}{k} (s_{01} + s_{10})^k, $$
we get 
\begin{equation}
S(M\tau) = 1 - Ms_{01}+s_{01} \sum_{k = 1}^{M-1} (-1)^{k+1} \binom{M}{k+1} (s_{01} + s_{10})^{k}.
\label{result2}
\end{equation}
We can perform simple checks on our results. We first set $s_{10} = 0$. Then it is obvious that $S(M\tau) = s_{00}^M$ in this case - once the system makes a transition to the state $\rho_1$, it cannot make a transition back to $\rho_0$. Eq.~\eqref{result1} reproduces this result, and, using $\sum_{k = 1}^{M - 1} (-1)^{k + 1} \binom{M}{M + 1} s_{01}^k = \frac{Ms_{01} + (1 - s_{01})^M - 1}{s_{01}}$, so does Eq.~\eqref{result2}. Furthermore, for $M = 2$, it is obvious that we should get $S = s_{00}^2 + s_{01}s_{10} = 1 - 2s_{01} +s_{01}(s_{01} + s_{10})$. One can check that we get the same result using Eqs.~\eqref{result1} and \eqref{result2}.

Let us now consider non-selective measurements where, after every time interval $\tau$, we perform a projective measurement on the system as before, but now we do not read the measurement results. We know from measurement theory that if the state just before the measurement is $\rho$, then the state just after the measurement is  $\rho' = \sum_i P_i \rho P_i$, where $P_i$ are the projection operators onto the eigenstates of the observable being measured \cite{vonNeumannbook,Wisemanbook}. It follows that if the initial state is $\rho_0$, the system state just after the first non-selective measurement is $\sum_{i_1} s_{0 i_1} \rho_{i_1}$. The state just after the second non-selective measurement is $\sum_{i_1 i_2} s_{0 i_1} s_{i_1 i_2} \rho_{i_2}$. Similarly, just after $M - 1$ non-selective measurements, the state of the system is $\sum_{i_1 i_2 \hdots i_{M - 1}} s_{0 i_1} s_{i_1 i_2} s_{i_2 i_3} \hdots s_{i_{M - 2} i_{M - 1}} \rho_{i_{M - 1}}$. The probability that a final selective measurement leads to $\rho_0$ is then 
$$ S(M\tau) =   \sum_{i_1 i_2 \hdots i_{M - 1}} s_{0i_1}s_{i_1 i_2}s_{i_2 i_3} \hdots s_{i_{M - 2}i_{M - 1}} s_{i_{M - 1}0}, $$
which is the same as Eq.~\eqref{finalprobability}. Thus, if we do not read off the measurement results, we obtain exactly the same results as before for the effective decay rate. Whether or not we read the measurement results makes no difference. 

We now illustrate the effect of repeated non-selective measurements using our formalism. Before doing so however, it is useful to note that Eq.~\eqref{result2} shows the dependence of the total survival probability on the system-environment coupling strength in a very transparent manner. Suppose that the system-environment coupling is very weak. Then $s_{01}$ and $s_{10}$ are very small, and can be calculated using first order time-dependent perturbation theory \cite{Chaudhryscirep2016}. It follows that $S(M\tau) \approx 1 - Ms_{01}$, which corresponds to $\Gamma(\tau) = s_{01}/\tau$. This is the usual result for the decay rate in the weak coupling regime. Thus considering non-selective measurements only has an effect on the total survival probability, and hence the effective decay rate, if we go beyond simple first-order perturbation theory. Consequently, we now illustrate the effect of considering unsuccessful measurements using exactly solvable models where we can calculate $s_{01}$ and $s_{10}$ exactly in regimes beyond the applicability of first order perturbation theory.

\subsection*{Single spin pure dephasing model}

We first study a single spin-$1/2$ particle interacting with an environment of harmonic oscillators. The total system-environment Hamiltonian is (we set $\hbar = 1$ throughout) \cite{BPbook}
\begin{equation}\label{p2}
H = \frac{\omega_0}{2}\sigma_z + \sum_k \omega_k b_k^\dagger b_k + \frac{\sigma_z}{2} \sum_k (g_k^* b_k + g_k b_k^\dagger),
\end{equation}
where the system Hamiltonian is $H_{S}=\frac{\omega_0}{2}\sigma_z$, the environment Hamiltonian is $H_{B} =\sum_k \omega_k b_k^\dagger b_k$,
while system-environment interaction Hamiltonian is $H_{SB} = \frac{\sigma_z}{2} \sum_k (g_k^* b_k + g_k b_k^\dagger)$.
Here $\omega_0$ is the energy spacing of two-level system, and $\omega_k$ denote the frequencies of the harmonic oscillator, while $b_k$ and $b^{\dagger}_k$ are the annihilation and creation operators for the harmonic oscillators, with $g_k$ is the coupling strength between the central spin system and the environment oscillators. An important feature of this model is that only the off-diagonal elements of the system density matrix (in the $\sigma_z$ eigenbasis) change in time, which is why this model is referred to as the pure dephasing model. 

Consider the initial state of the two-level system to be $\ket{\psi_0} = \cos\left(\frac{\theta}{2}\right)\ket{e} + e^{i\phi}\sin\left(\frac{\theta}{2}\right)\ket{g}$ with $\ip{e}{g} = 0$. The states $\ket{g}(\ket{e}) $ are the ground (excited) states of the spin-1/2 particle, and $\theta$ and $\phi$ are parameters characterizing the state $\ket{\psi_0}$. The state orthogonal to this state is $\ket{\psi_1} = \sin\left(\frac{\theta}{2}\right)\ket{e} - e^{i\phi}\cos\left(\frac{\theta}{2}\right)\ket{g}$. At time intervals $\tau$, we perform non-selective measurements in the basis $\lbrace \ket{\psi_0},\ket{\psi_1}\rbrace$. If the state of the system is $\rho_0 = \ket{\psi_0}\bra{\psi_0}$, the probability that the system ends up in state $\rho_1 = \ket{\psi_1}\bra{\psi_1}$ a time interval $\tau$ later (after removal of the evolution due to the system Hamiltonian) is (see the Methods section)
\begin{equation*}
s_{01} = \frac{1}{2}\sin^2\theta (1 - e^{-\gamma(\tau)}).
\end{equation*}
Here $\gamma(\tau)=\sum_k \frac{|g_k|^2}{\omega_k^2}[1-\cos(\omega_k\tau)]\coth(\beta\omega_k/2)$ describes the environment-induced dephasing, that is the loss of coherence between the states $\ket{e} $ and $\ket{g}$. To perform the sum over the oscillator modes, we will replace the sum by an integral via $\sum_k |g_k|^2 (\hdots) \rightarrow \int_0^\infty \, d\omega \, J(\omega) (\hdots)$ \cite{BPbook}. Throughout, we will use an Ohmic spectral density with an exponential cutoff to illustrate our results, that is, $J(\omega) = G \omega e^{-\omega/\omega_c}$, where $G$ is the dimensionless system-environment coupling strength, and $\omega_c$ is the cutoff frequency. We have also assumed that the initial system-environment state is $\rho_0 \otimes \rho_B$, with $\rho_B = e^{-\beta H_B}/Z_B$ and $Z_B = \tr[e^{-\beta H_B}]$. 

\begin{figure}[t]
	\includegraphics[scale = 0.7]{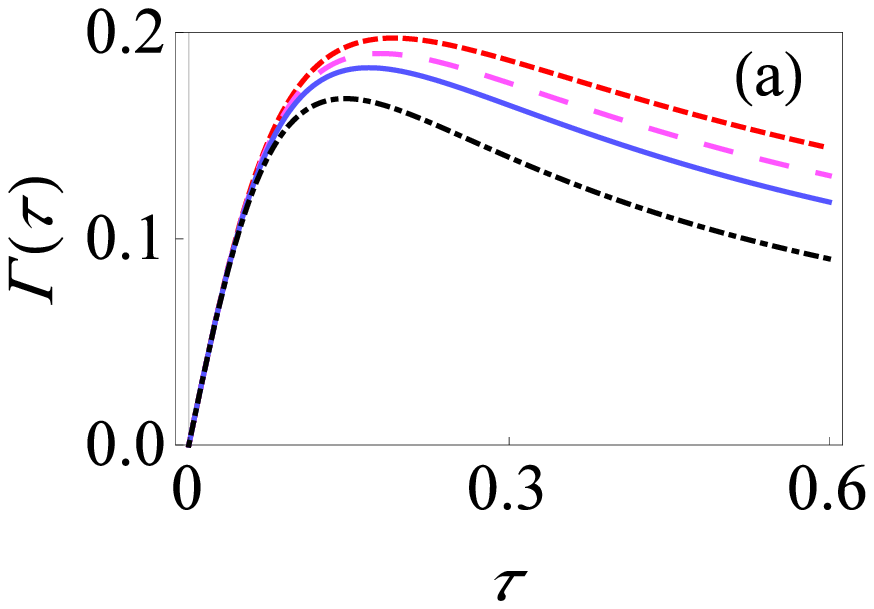}
	\qquad
	\qquad
	\includegraphics[scale = 0.7]{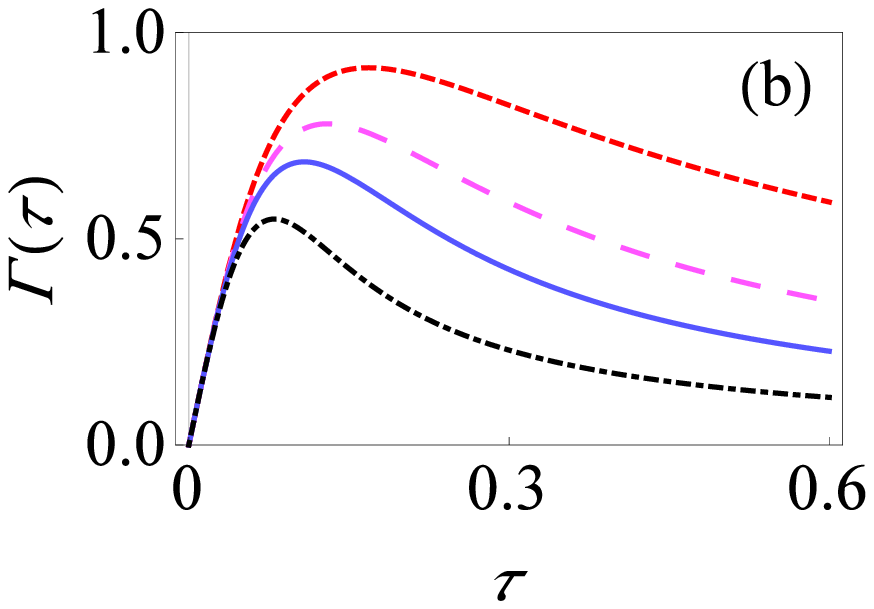}
	\centering
	\caption{ \textbf{Effective decay rate for single spin pure dephasing model}. $\textbf{(a)}$  Behavior of $\Gamma(\tau)$ versus $\tau$ for the initial state of the central spin $\psi_0$ with only selective measurements (small-dashed, red curve),  and with non-selective measurements with $M=3$ (large-dashed, magenta curve),  $M=5$ (solid, blue curve) and  $M=10$ (dot-dashed, black curve). We work in dimensionless units with $\hbar= 1$. Here we have set $\theta=\pi/2$, $\phi=0 $, $\omega_0=1$, $\beta=10$, $\omega_c=10$ and $G=0.1$. $\textbf{(b)}$ Same as $\textbf{(a)}$, except that now $G=0.5$.  }
	\label{SSH}
\end{figure}

In a similar manner, we find that if the system state is $\rho_1$, the probability that after time interval $\tau$ the system state is found to be $\rho_0$ is 
\begin{equation*}
s_{10} = \frac{1}{2}\sin^2\theta (1 - e^{-\gamma(\tau)}).
\end{equation*}
Thus, in this case, the transition probabilities are the same. Let us denote $s_{01} = s_{10} = s$. Using Eq.~\eqref{result1} gives the following form of survival probability
\begin{equation}
S(M\tau) = \frac{1}{2}\left[1 + (1 - 2s)^M\right].
\end{equation} 
The corresponding effective decay rate is 
\begin{equation}\label{SSM3}
\Gamma(\tau) = -\frac{1}{M\tau} \ln \left\lbrace \frac{1}{2}\left[1 + [1 - \sin^2 \theta(1 - e^{-\gamma(\tau)})]^M\right]\right\rbrace.
\end{equation}
This expression should be compared with that obtained by performing only selective measurements. In the latter case, we simply have 
\begin{equation}
\Gamma(\tau) = -\frac{1}{\tau} \ln [ 1 - \frac{1}{2}\sin^2 \theta (1 - e^{-\gamma(\tau)})].
\end{equation} 
In Fig.~\ref{SSH}, we show the behavior of the decay rate $\Gamma(\tau)$ as a function of measurement interval $\tau$ with weak [Fig.~\ref{SSH}(a)] and relatively strong [Fig.~\ref{SSH}(b)] system-environment coupling strength at low temperatures. It is clear that we observe both the quantum Zeno and anti-Zeno regimes. For smaller values of $\tau$, the effective decay rate $\Gamma(\tau)$ decreases as the measurement interval $\tau$ is reduced, meaning that shorter measurement interval $\tau$ helps to protect the state of quantum system, thus putting us in quantum Zeno regime. However for larger values of $\tau$, the opposite situation takes place, namely, effective decay rate increases as the $\tau$ decreases, hence indicating the anti-Zeno regime for both selective and non-selective measurements. Furthermore, especially with relatively strong system-environment coupling, only three measurements can bring out a significant difference between performing non-selective measurements and performing only selective measurements (compare the small-dashed, red curve with the large-dashed, magenta curve). We notice that as we increase the number of non-selective measurements, the effective decay rate reduces. The value of $\tau$ for which we make a transition from the Zeno regime to the anti-Zeno regime also shifts to a lower value. These trends become more prominent with stronger system-environment coupling [compare Figs.~\ref{SSH}(a) and (b)].

\subsection*{Spin interacting with spin environment}

We now consider a single spin-$1/2$ particle interacting with an environment of $N$ other spin-1/2 particles. Our total system-environment Hamiltonian is \cite{cucchietti2005decoherence, Chaudhryarxiv2018}
\begin{equation}\label{p1}
H = \dfrac{\varepsilon}{2}\sigma_{z} + \dfrac{\Delta}{2}\sigma_{x} + \sum_{i=1}^{N}\dfrac{\varepsilon_{i}}{2}\sigma_{z}^{(i)}+ \frac{\sigma_{z}}{2}\otimes\sum_{i=1}^{N}g_{i}\sigma_{z}^{(i)},
\end{equation}
where the first term is the central spin Hamiltonian is $H_{S}$, the environment Hamiltonian $H_{B}$ is given by the second term, while the third term describes the the system-environment interaction $H_{SB}$. Here $\Delta$ and $\varepsilon$ denote the tunneling amplitude and the energy spacing of the central spin system respectively, $\sigma_m$ $(m=x,y,z)$ are the standard Pauli spin operators as before, $\varepsilon_{i}$ is the energy spacing in case of  $i^{\text{th}}$ environmental spin, and $g_i$ describes the interaction strength between the central spin system and the $i^{\text{th}}$ environmental spin. An important feature of this model is that now both the diagonal and off-diagonal elements of the central spin density matrix change with time. The dynamics given by this model can also be solved exactly with the initial system environment state given by $\rho_{\text{tot}}(0) = \rho_S(0) \otimes \rho_B$ where $\rho_B = e^{-\beta H_B}/Z_B$ is the thermal equilibrium state of the environment. We defer the details to the Methods section, but it is pertinent to note here that the key to solving this system-environment Hamiltonian is that the environment Hamiltonian $H_B$ commutes with the environment part of the system-environment interaction Hamiltonian. The joint eigenstates can be written as $\ket{n} = \ket{n_1} \ket{n_2} \hdots \ket{n_N}$, with $n_i = 0 (1)$ denoting the spin up (down) state (along the $z$ direction). The initial state of the central spin that we choose is $\rho_0 = \frac{1}{2}(1 + \sigma_x)$. Correspondingly, $\rho_1 = \frac{1}{2}(1 - \sigma_x)$. The probability that, starting from the state $\rho_0$, after time $\tau$ we find the state $\rho_1$ is given by (see the Methods section)
 \begin{equation}
 \label{probspinenv}
s_{01}(\tau) = \frac{1}{2}\left[1 - p_x(\tau)n_x(\tau) - p_y(\tau)   n_y(\tau)  - p_z(\tau)  n_z(\tau)\right] , 
\end{equation}
where 
\begin{align}
p_x{(\tau)} =  \frac{1}{Z_B}\sum_{n} \frac{c_n}{4\Omega_n^2} (\zeta_n^2 \cos(2\Omega_n t) + \Delta^2 ), \; p_y{(\tau)} =  \frac{1}{Z_B}\sum_{n} \frac{c_n}{2\Omega_n} \zeta_n \sin(2\Omega_n t), \; p_z{(\tau)} =  \frac{1}{Z_B}\sum_{n} \frac{c_n}{2\Omega_n^2} \Delta \zeta_n \sin^2(\Omega_n t), 
\label{spinenvtransprob}
\end{align}
and 
\begin{align}
n_x{(\tau)} = [\cos^2 (\Omega \tau) + \frac{\sin^2(\Omega \tau)}{4\Omega^2} (\Delta^2 - \varepsilon^2)],\; n_y{(\tau)} = \frac{\varepsilon}{\Omega} \sin(\Omega \tau) \cos(\Omega \tau), \; n_z{(\tau)} =  \frac{\varepsilon\Delta}{2} \frac{\sin^2(\Omega \tau)}{\Omega^2}.
\label{spinenvnprob}
\end{align}
Here $c_n = e^{-\beta \eta_n/2}$ with $\eta_n = \sum_{i = 1}^N (-1)^{n_i} \varepsilon_i$, $Z_B = \sum_n c_n$, $\zeta_n = \varepsilon + G_n$ with $G_n = \sum_{i = 1}^N (-1)^{n_i} g_i$, $\Omega_n^2 = \frac{1}{4}(\zeta_n^2 + \Delta^2)$ and $\Omega^2 = \frac{1}{4}(\epsilon^2 + \Delta^2)$. We also find that $s_{01} = s_{10}$. Consequently, denoting $s_{01} = s$ and using Eq.~\eqref{result1}, we get
\begin{equation*}\label{SP}
S(M\tau) =   \frac{1}{2}[1+(1 - 2s)^M] ,
\end{equation*}
leading to the effective decay rate
\begin{equation}\label{SSM1}
\Gamma(\tau) = -\frac{1}{M\tau} \ln\{ \frac{1}{2} [1 + (p_x(\tau)n_{x}(\tau)+p_y(\tau)n_{y}(\tau)+p_z(\tau)n_{z}(\tau))^M ]\}.
\end{equation}
This result should be compared with repeated selective measurements where the effective decay rate is independent of number of measurements and has the form
\begin{equation*}\label{SSM2}
\Gamma(\tau) = -\frac{1}{\tau} \ln\{ \frac{1}{2} [1 + (p_x(\tau)n_{x}(\tau)+p_y(\tau)n_{y}(\tau)+p_z(\tau)n_{z}(\tau))]\}.
\end{equation*}

\begin{figure}[t!]
	\includegraphics[scale = 0.7]{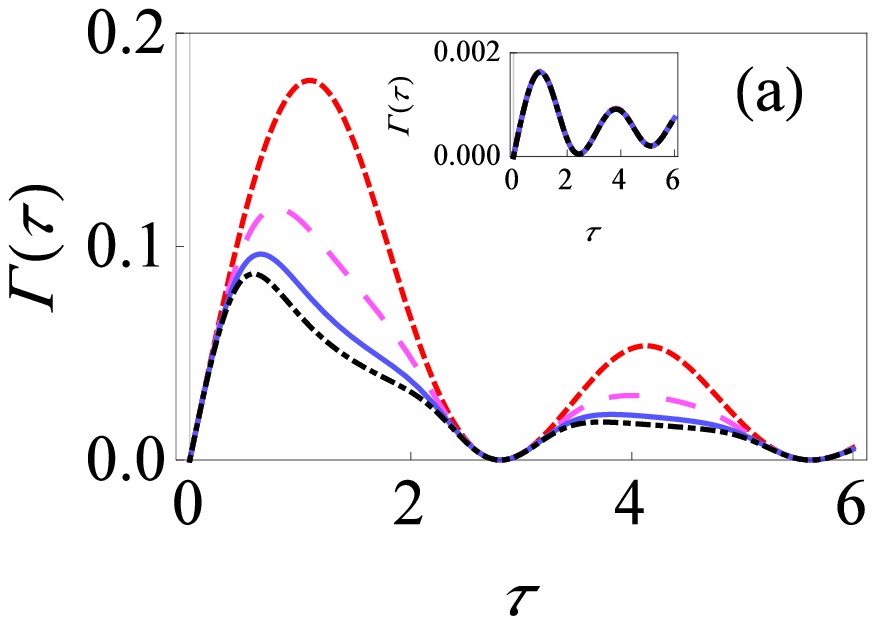}
    \qquad
     \qquad
	\includegraphics[scale = 0.7]{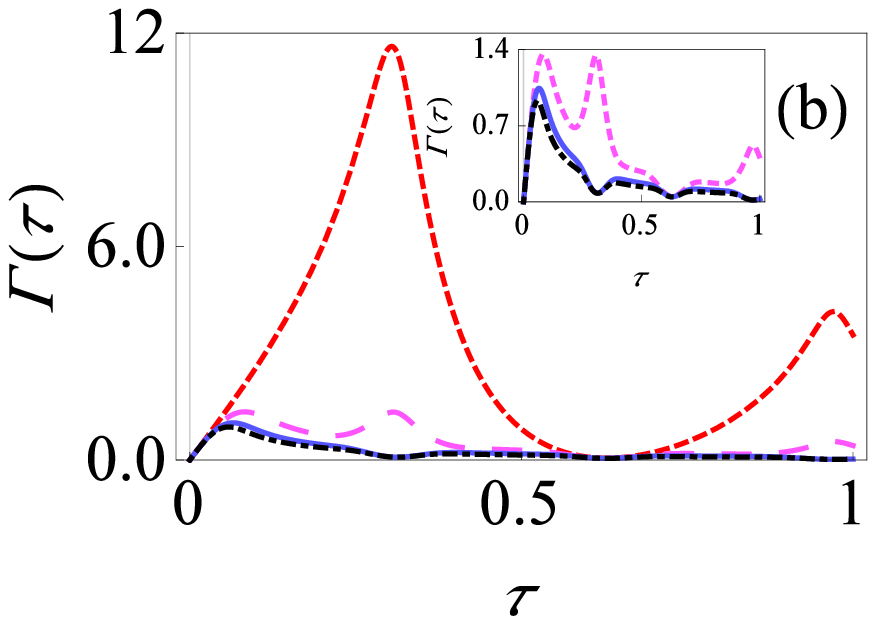}
	\caption{ \textbf{ Effective decay rate for the spin-spin environment model}. \textbf{(a)}  Behavior of $\Gamma(\tau)$ versus $\tau$ for the initial state of central system prepared along $ \ket{\uparrow_{x}}$ with $M=1$ (small-dashed, red curve), $M=3$ (large-dashed, magenta curve), $M=5$ (solid, blue curve) and  $M=10$ (dot-dashed, black curve) for the spin environment. We work in dimensionless units so that $\hbar= 1$. For simplicity, level spacing $\epsilon_i$ and coupling strength $g_i$ are chosen to have the same value for every environment. Here we have set $\epsilon=1$, $\Delta=2$, $\beta=10$, $\epsilon_i=1$, $g_i=0.01$ and the number of environmental spins is considered to be $N=100$. The inset shows the effective decay rate with the same system-environment parameters, except that now $g_i=0.001$). \textbf{(b)} Same as \textbf{(a)}, except that now $g_i=0.1$, with zoomed-up inset plot.}
	\label{SSE}
\end{figure}

In Fig.~\ref{SSE}, the effective decay rate $\Gamma(\tau)$ with the spin environment has been plotted as a function of the measurement interval $\tau$  for different values of system-environment parameters, again at very low temperatures. The small-dashed red curve is the decay rate if we perform only selective measurements, while with non-selective measurements, the large-dashed magenta curve is the decay rate for $M=3$ , solid blue curve is the decay rate for $M=5$ and dot-dashed black curve is the decay rate for $M=10$. Let us first focus on the inset of Fig.~\ref{SSE}. As mentioned before, for very weak system-environment coupling, $s_{01}$ and $s_{10}$  approach to zero; consequently, the effective decay rate will be $\Gamma(\tau)\approx-\frac{1}{\tau}s_{10}$, independent of the number of measurements. Thus, both selective and non-selective measurements lead to the same effective decay rate in such a case, independent of the number of measurements. This is precisely the case in the inset where the curves overlap. However, with stronger system-environment coupling strength, higher order terms in Eq.~\eqref{result2} also contribute, making the effective decay rate different for the selective and non-selective cases. This is illustrated in the main figure of Fig.~\ref{SSE}(a). With both selective and non-selective measurements, there exist distinct multiple quantum Zeno and anti-Zeno regimes, that is, sometimes decay rate decreases by decreasing the measurements interval $\tau$ (meaning that we are in the quantum Zeno regime), while sometimes it increases by decreasing the $\tau$ (meaning that we are in the anti-Zeno regime). For repeated non-selective measurements, we clearly see that once again the effective decay rate $\Gamma(\tau)$ is lower compared to only selective measurements, and the decay rate further reduces as the number of measurements is increased. Moreover, as before in our study of the single spin pure dephasing model, the peak value of the decay rate is shifted to the smaller values of $\tau$. With even stronger system-environment coupling strength, these differences become even more pronounced, as illustrated in Fig.~\ref{SSE}(b). With selective measurements, multiple quantum Zeno to anti-Zeno transitions exist, but these transitions are less with non-selective measurements due to the smaller values of effective decay rates. Consequently, the differences in the effective decay rates translate to very significant differences in the final survival probabilities.

\subsection*{Large spin pure dephasing model}

\begin{figure}[t]
 	\includegraphics[scale = 0.7]{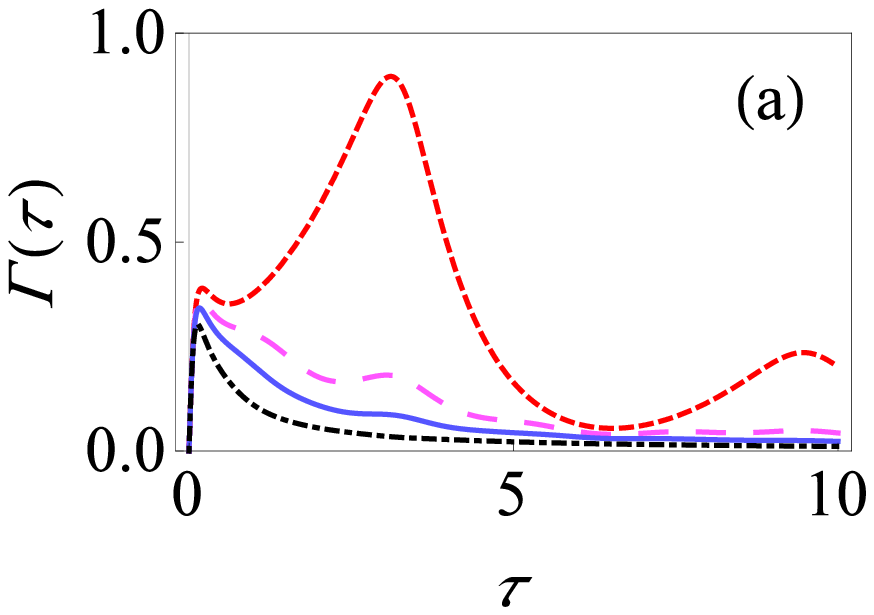}
 	\qquad
 	\quad
 	\includegraphics[scale = 0.7]{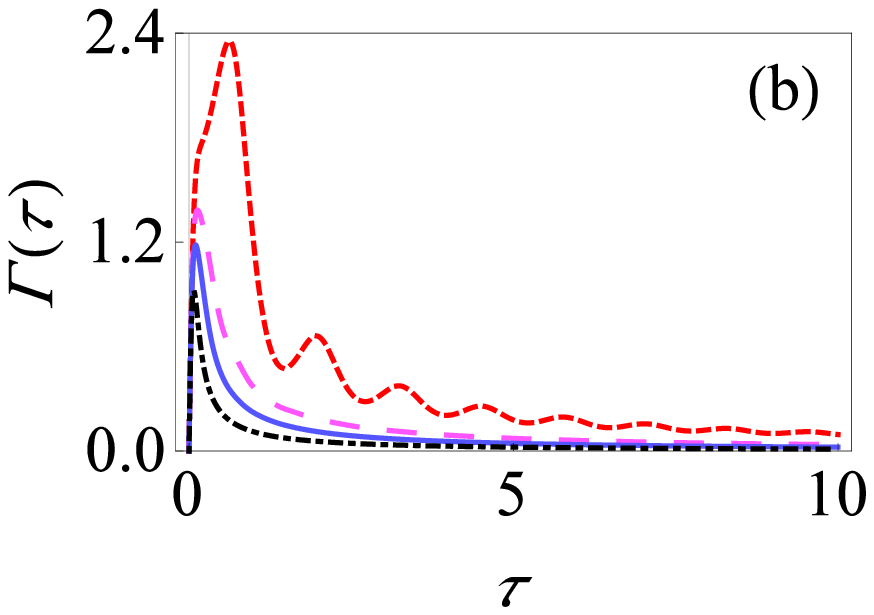}
 	\centering	
 	\caption{ \textbf{ Effective decay rate for the large spin pure dephasing model with $J=1$}. $\textbf{(a)}$  Behavior of $\Gamma(\tau)$ versus $\tau$ for the initial state $\rho_0$ (spin up in the $x$ direction) with only selective measurements (small-dashed, red curve), and using non-selective measurements with $M=3$ (large-dashed, magenta curve),  $M=5$ (solid, blue curve) and  $M=10$ (dot-dashed, black curve).  Here we have set $\omega_0=1$, $\beta=10$, $\omega_c=10$ and $G=0.1$. $\textbf{(b)}$ Same as $\textbf{(a)}$, except that now $G=0.5$.  }
 	\label{LSHO}
 \end{figure}

To further illustrate our formalism, we now consider a scenario beyond a simple two-level system. We consider in particular a spin $J = 1$ particle interacting with harmonic oscillator environment. Such a model can describe the physics of two spin-1/2 particles interacting a common harmonic oscillator environment. The system-environment Hamiltonian is now 
\begin{equation}
\label{largespinHamiltonian}
H = \omega_0 J_z + \sum_k \omega_k b_k^\dagger b_k + J_z \sum_k (g_k^* b_k + g_k b_k^\dagger),
\end{equation}
where $J_z$ is is the usual angular momentum operator and the remaining parameters are described as before. For the simplicity of presentation, let us suppose that we repeatedly measure the operator $J_x$. The initial system state that we prepare is the eigenstate of $J_x$, with eigenvalue $+1$. Written in the standard $J_z$ eigenbasis, this state is 
$$ \rho_0 = \frac{1}{4} \left( \begin{array}{ccc}1 & \sqrt{2} & 1 \\ \sqrt{2} & 2 & \sqrt{2} \\ 1 & \sqrt{2} & 1 \end{array}\right). $$
The other two orthogonal eigenstates of $J_x$ are 
 $$ \rho_1 = \frac{1}{2} \left( \begin{array}{ccc}1 & 0 & -1 \\ 0 & 0 & 0 \\ -1 & 0 & 1 \end{array}\right), $$
 and 
 $$ \rho_2 = \frac{1}{4} \left( \begin{array}{ccc}1 & -\sqrt{2} & 1 \\ -\sqrt{2} & 2 & -\sqrt{2} \\ 1 & -\sqrt{2} & 1 \end{array}\right). $$
 Knowing the Hamiltonian, we can work out the system density matrix at any time exactly. Assuming that the initial system-environment state is $\rho_S(0) \otimes e^{-\beta H_B}/Z_B$, the result, written in the $J_z$ eigenbasis after the removal of the evolution due to $H_S$, is (see the Methods section)
 $$ [\rho_S(t)]_{lm} = [\rho_S(0)]_{lm} e^{-i\delta(t)(l^2 - m^2)}e^{-\gamma(t)(l - m)^2}. $$
 Here $\gamma(\tau)$ is the decoherence factor defined before, and $\delta(\tau) =\sum_k \mid g_k\mid^2(\sin(\omega_k \tau)-\omega_k\tau)/\omega_k^2$ describes the indirect interaction between the two two-level systems due to the common environment. It is then simple to work out that  
 \begin{equation}
 s_{01} = \frac{1}{4} \left[1 - e^{-4\gamma(\tau)}\right] = s_{10} = s_{12} = s_{21}, 
 \end{equation}
 and 
 \begin{equation}
 s_{02} = \frac{1}{8} \left[3 + e^{-4\gamma(\tau)} - 4 \cos[\delta(\tau)] e^{-\gamma(\tau)}\right] = s_{20}.
 \end{equation}
 Our objective to now evaluate Eq.~\eqref{finalprobability} in this case. The result is (see the Methods section)
 \begin{equation}
 \label{problargespin}
 S(M\tau) = \frac{1}{6} \left[ 2 + (1 - 3s_{01})^M + 3(1 - s_{01} - 2s_{02})^M\right],
 \end{equation} 
 and the corresponding effective decay rate is
  \begin{equation}\label{decayratelargespin_boson}
 \Gamma(\tau) =-\frac{1}{M \tau}\ln\left\{\frac{1}{6} \left[ 2 +\dfrac{1}{4^M} (1 + 3e^{-4\gamma(\tau)})^M+ 3(\cos(\delta(\tau)) e^{-\gamma(\tau)})^M\right]\right\}. 
 \end{equation}
In contrast, for selective measurements, the effective decay rate is
  \begin{equation*}\label{decayratelargespin_bosonsuccessful}
 \Gamma(\tau) =-\frac{1}{ \tau}\ln\left\{\frac{1}{8} \left[ 3 +4 e^{-4\gamma(\tau)}+ 4\cos[\delta(\tau)] e^{-\gamma(\tau)}\right]\right\} . 
 \end{equation*}
The key difference now as compared to the single spin pure dephasing model is the presence of the $\delta(\tau)$ term that describes the effect of the indirect interaction between the spins. In Fig.~\ref{LSHO}, we illustrate the behavior of the effective decay rate $\Gamma(\tau)$ as a function of the measurement interval $\tau$. If we perform selective measurements with relatively weak system-environment coupling strength, it is clear that we observe distinct Zeno and anti-Zeno regimes [see Fig.~\ref{LSHO}(a)]. Comparing with the single spin case, we note that the indirect interaction between the spins is responsible for the multiple Zeno and anti-Zeno transitions. However, with non-selective measurements, we largely observe one Zeno regime and one anti-Zeno regime. This is because, as before, the non-selective measurements lead to a lowering of the effective decay rate and the measurement interval at which the peak effective decay rate occurs shifts to lower values as well. However, for smaller values of $\tau$, the indirect interaction plays a relatively smaller role - it can be checked that $\delta(\tau) \rightarrow 0$ as $\tau \rightarrow 0$. On the other hand, for stronger system-environment coupling strength, as illustrated in Fig.~\ref{LSHO}(b), the decoherence factor $\gamma(\tau)$ plays a more dominant role as compared to the indirect interaction $\delta(\tau)$. Consequently, there are now less clear cut multiple Zeno and anti-Zeno regimes.

 \section*{Discussion}
 
 In this paper, we have generalized the treatment of the quantum Zeno and anti-Zeno effects by considering non-selective measurements. We have worked out a general formalism for calculating the effective decay rate of a quantum state subjected to repeated non-selective measurements. Importantly, we have shown that non-selective measurements lead to a different effective decay rate as compared to the usual strategy of using only selective measurements if we go beyond the weak system-environment coupling regime. To illustrate our formalism, we also worked out the effective decay rate for three exactly solvable system-environment models. These included a single spin interacting with a harmonic oscillator environment, a single spin interacting with a spin environment, and two spins interacting with a harmonic oscillator environment. Using these exactly solvable models, we found that non-selective measurements can qualitatively alter the analysis of the quantum Zeno and anti-Zeno effects. In particular, non-selective measurements  considerably reduce the effective decay rate, and the transition from Zeno to anti-Zeno regimes (and vice versa) is also altered. Our results should be important in the study of the effect of repeated measurements if we go beyond the weak system-environment coupling regime.

\section*{Methods}

\subsection*{Finding the final survival probability}
To evaluate Eq.~\eqref{finalprobability}, we can employ matrix multiplication. Define the matrix $\mathcal{S}$ as 
$$ \mathcal{S} = 
  \left( {\begin{array}{cc}
   1 - s_{01} & s_{01} \\
   s_{10} & 1 - s_{10} \\
  \end{array} } \right). $$
Then it is straightforward to see that $S(M\tau) = [\mathcal{S}^M]_{00}$, that is, $S(M\tau)$ is simply the top-left element of the matrix $\mathcal{S}^M$. Our problem is then to $\mathcal{S}^M$. This can be done via diagonalization. Define $D = U^{-1} \mathcal{S} U$, where $D$ is a diagonal matrix with the eigenvalues of $\mathcal{S}$ as its diagonal elements and $U$ is a matrix with eigenvectors of $\mathcal{S} $ as its columns. Then,
$$ D = 
\left( {\begin{array}{cc}
	1  & 0 \\
	0 & 1 - s_{01}-s_{10} \\
	\end{array} } \right), \; 
	U = \left( {\begin{array}{cc}
	1  & -\frac{s_{01}}{s_{10}} \\
	1 & 1  \\
	\end{array} } \right), \; 
	U^{-1} = \frac{1}{s_{01} + s_{10}} \left( {\begin{array}{cc}
	s_{10}  & s_{01} \\
	-s_{10} & s_{10}  \\
	\end{array} } \right),$$
and $\mathcal{S}^M = U D^M U^{-1}$ is 
$$ \mathcal{S}^M = 
\left( {\begin{array}{cc}
	\dfrac{s_{01}(1-s_{01}-s_{10})^{M}+s_{10}}{s_{01}+s_{10}}  & \dfrac{s_{01}-s_{01}(1-s_{01}-s_{10})^{M}}{s_{01}+s_{01}} \\
	\dfrac{s_{10}-s_{10}(1-s_{01}-s_{10})^{M}}{s_{01}+s_{01}} & \dfrac{s_{01}+s_{10}(1-s_{01}-s_{10})^{M}}{s_{01}+s_{01}} \\
	\end{array} } \right). $$
Consequently, we can read off that $S(M\tau)$ is as given in Eq.~\eqref{result1}.

A very similar method can be employed for a higher dimensional system. Consider, for example, a three dimensional systems as is the case for the large spin pure dephasing model. In this case, we construct  
$$ \mathcal{S} = 
  \left( {\begin{array}{ccc}
   1 - s_{01} - s_{02} & s_{01} & s_{02} \\
   s_{10} & 1 - s_{10} - s_{12} & s_{12}\\
   s_{20} & s_{21} & 1 - s_{20} - s_{21}
  \end{array} } \right). $$
Then, once again, $S(M\tau)$ is simply the top-left element of the matrix $\mathcal{S}^M$. Again, the task is to simply diagonalize $\mathcal{S}$. However, in this case, the algebra is much more cumbersome for the general case. Fortunately, for the pure dephasing model, $s_{01} = s_{10} = s_{12} = s_{21}$, and $s_{20} = s_{02}$, which leads to great simplifications. In this case, following the same method as above,  
$$ D = 
\left( {\begin{array}{ccc}
	1  & 0 & 0\\
	0 & 1 - 3s_{01} & 0 \\
	0 & 0 & 1 - s_{01} - 2s_{02}
	\end{array} } \right), \; 
	U = \left( {\begin{array}{ccc}
	1  & 1 & -1 \\
	1 & -2 & 0  \\
	1 & 1 & 1
	\end{array} } \right), \; 
	U^{-1} = \frac{1}{6} \left( {\begin{array}{ccc}
	2  & 2 & 2 \\
	1 & -1 & 2  \\
	-3 & 0 & 3
	\end{array} } \right),$$
and the top left element of $\mathcal{S}^M = U D^M U^{-1}$ is then given by Eq.~\eqref{problargespin}.
	
\subsection*{Derivation of the spin density matrix with harmonic oscillator environment}
Let us now, for completeness, outline how to find the system density matrix with the system-environment Hamiltonian given in Eq.~\eqref{largespinHamiltonian}. Further details can be found, for example, in Ref.~{\renewcommand{\citemid}{}\cite[]{ChaudhryPRA2013a}. The single spin density matrix can then be found by simply setting the spin size to $1/2$. Our first goal is to find the  total unitary time-evolution operator $U(\tau)$. To this end, it is useful to first write $U(\tau)=U_{F}(\tau)U_I(\tau)$, where  $U_{F}(\tau)=e^{-i(H_{S}+H_B)\tau}$ is the free unitary time time evolution operator and $U_I(\tau)$ is the time evolution due to the system-environment interaction. One can then show, using the Magnus expansion, that $U_I(\tau)=\text{exp}[J_z\sum_k(b^{\dagger}_k\alpha_k(\tau)-b_k \alpha^*_k(\tau))-iJ_z^2\delta(t)]$, where $\alpha_k(\tau)=g_k(1-e^{i{\omega_k}\tau})/\omega_k$ and $\delta(\tau)=\sum_k |g_k|^2(\sin(\omega_k \tau)-\omega_k\tau)/\omega_k^2$. With the time evolution operator found, we can write the system density matrix $\rho_S(\tau)$ in terms of $J_z$ eigenbasis as $[\rho_S(\tau)]_{lm}=\tr_{S,B}[U(\tau)\rho_{\text{tot}}(0)U^{\dagger}(\tau)P_{lm}]$. Here $P_{lm}=\ket{l}\bra{m}$, where $\ket{l} $ is the eigenstate of operator $J_z$  with eigenvalue $l$. Assuming an initially uncorrelated system-environment state with the environment in thermal equilibrium, that is, $\rho_{\text{tot}}(0)=\rho_S(0)\otimes\rho_B$ with $\rho_B = e^{-\beta H_B}/Z_B$ and $Z_B = \tr[e^{-\beta H_B}]$, we obtain $ [\rho_S(\tau)]_{lm}=e^{-i\omega_{0}\tau(l-m)}e^{-i\delta(\tau)(l^2-m^2)}[\rho_S(0)]_{lm}\av{e^{-R_{lm}(\tau)}}$, with $R_{lm}(\tau) = (l - m) \sum_k [b_k^\dagger \alpha_k(\tau) - b_k \alpha_k^*(\tau)]$, and $\tr_{B}[e^{-R_{lm}(\tau)}\rho_B]=\av{e^{-R_{lm}(\tau)}}$ is the average over the thermal states of the bath in equilibrium. This average is found to be $\tr_{B}[e^{-R_{lm}(\tau)}\rho_B]=\text{exp}[-\sum_k(l-m)^2 |g_k|^2(1-\cos(\omega_k\tau))\coth(\beta\omega_k/2)/\omega_k^2] $. Consequently, all in all, we have 
$$[\rho_S(\tau)]_{lm}=e^{-i\omega_{0}\tau(l-m)}e^{-i\delta(\tau)(l^2-m^2)}[\rho_S(0)]_{lm}e^{-\gamma(\tau)(l-m)^2},$$ 
with $\gamma(\tau)=\sum_k\mid g_k\mid^2(1-\cos(\omega_k\tau))\coth(\beta\omega_k/2)/\omega_k^2$.

We are really interested in finding the transition probabilities. Suppose that the initial system state is $\rho_0$. Then the probability that a measurement at time $\tau$ yields the state $\rho_1$ (after removal of the evolution due to $H_S$) is 
$$ s_{01} = \sum_{lm} e^{-i\delta(\tau)(l^2-m^2)}e^{-\gamma(\tau)(l-m)^2} [\rho_0]_{lm} [\rho_1]_{ml}. $$
Other survival probabilities can be calculated in an analogous manner. 

\subsection*{Solving the central spin-spin environment model}
We now outline how to find the spin dynamics with system-environment model given in Eq.~\eqref{p1}. Details can be found in Ref.~{\renewcommand{\citemid}{}\cite[]{Chaudhryarxiv2018}. We first write the interaction term between the central system and the environment as $H_{\text{SB}} = \frac{1}{2}\sigma_z \otimes B$, where  $B$ is the environment operator, defined to be $B = \sum_{i = 1}^N g_i\sigma_z^{(i)}$. The eigenstates of environment  operator $B$ can be expressed as the products of the eigenbasis $\ket{0_i}$ and $\ket{1_i}$ of $i^{\text{th}}$ environment spin operator ${\sigma}_{z}^{(i)}$, where $\ket{0}$ labels the spin `up'  and $\ket{1}$ the spin `down' state of the environment. As such, the eigenstates of environment operator $B$ can be written as $\ket{n} \equiv \ket{n_1} \ket{n_2} \hdots . \ket{n_N}$, with $n_i = 0, 1$. Explicitly ${B}\ket{n} = G_{n}\ket{n}$,  
with $G_{n}=\sum_{i=1}^{N}(-1)^{n_{i}}g_{i}$. Similarly,
$\sum_{i = 1}^N \frac{\varepsilon_i}{2} \sigma_z^{(i)}\ket{n} = \frac{1}{2}\eta_n\ket{n}$, with 
$\eta_{n} = \sum_{i=1}^{N}(-1)^{n_{i}}\varepsilon_{i}$. Since the environment states $\ket{n}\bra{n}$ commutes with the total Hamiltonian [see Eq. (\ref {p1})], we can find the unitary time-evolution operator for the total system by introducing the completeness relation over the environment states $\ket{n}$ i.e, $U(\tau) = \sum_{n=0}^{2^N-1} e^{-i\eta_n \tau/2} e^{-i(H_S + H_{SB})\tau} \ket{n}\bra{n},$
\begin{equation}
\label{unitarytimeoperator}
{U}(\tau)=\sum_{n=0}^{2^{N}-1}{U}_{n}(\tau)\ket{n}\bra{n},
\end{equation}
where $U_{n}(\tau)={e^{-i\eta_{n}t/2}}\left[\cos (\Omega_{n}\tau)-\dfrac{i}{\Omega_{n}}\sin(\Omega_{n}\tau)\left(\dfrac{\zeta_n}{2}\sigma_{z}+\dfrac{\Delta}{2}\sigma_{x}\right)\right]$, with $\zeta_n = \varepsilon + G_n$ and $\Omega_{n}^{2}=\frac{1}{4}\left(\zeta_n^2 + \Delta^2\right)$. For simplicity, we choose the initial system-environment state as $\rho_{\text{tot}}(0)=\rho_S(0) \otimes e^{-\beta H_B}/Z_B$, where $\rho_S(0) =  \rho_0=\frac{1}{2}\left(1 + \sigma_x\right)$ is the initial state of of central system, and $\rho_B=e^{-\beta H_B}/Z_B$ is the thermal equilibrium state of environment with $Z_B = \tr_B[e^{-\beta H_B}]$. Correspondingly, $\rho_1 = \frac{1}{2}\left(1 - \sigma_x\right)$ (the state orthogonal to initial state of system). Now the density matrix of central spin system  at some time $\tau$ is $\rho_S(\tau) = \tr_B[e^{-iH\tau}\rho_{\text{tot}}(0)e^{iH\tau}]$, where $\tr_{B}$ is the trace over the states of environment. Using Eq.~\eqref{unitarytimeoperator}, we get ${{\rho}_S}(\tau)= \frac{1}{Z_B}\sum_{n=0}^{2^{N}-1} c_n U_{n}(t)\rho_S (0)U_n^\dagger$, Here we have defined $c_n = {e}^{-{{\beta}\eta_{n}/2}}$, leading to $Z_{B}=\sum_n c_n$. Further simplification leads to 
$${{\rho}_S}(\tau)= \frac{1}{2} (1+p_x(\tau)\sigma_{x}+p_y(\tau)\sigma_{y}+p_z(\tau)\sigma_{z}), $$
where $p_x(\tau)$, $p_y(\tau)$ and $p_z(\tau)$ are defined in Eqs.~\eqref{spinenvtransprob}. The central spin system density matrix just before the measurement but after the removal of evolution due to free spin system Hamiltonian is then 
$${{\rho}_S}(\tau)= \frac{1}{2} (1+p_x(\tau)\sigma_{x}n_x(\tau)+p_y(\tau)\sigma_{y}n_y(\tau)+p_z(\tau)\sigma_{z}n_z(\tau)),$$
with $n_x(\tau)$, $n_y(\tau)$ and $n_z(\tau)$ defined in Eqs.~\eqref{spinenvnprob}. The probability $s_{01}$ can then be calculated in a straightforward manner leading to Eq.~\eqref{probspinenv}, while the calculation of $s_{10}$ is very similar and gives the same result.

\section*{Acknowledgements}

A.~Z.~C. and M.~M. acknowledge support from the LUMS FIF Grant FIF-413. A.~Z.~C. is also grateful for support from HEC under grant No 5917/Punjab/NRPU/R\&D/HEC/2016.

\section*{Author contributions statement}

A.~Z.~C. came up with the basic idea behind this work. M.~M.~ carried out the calculations and plotted the graphs. Both authors contributed towards the writing of the manuscript. 

\section*{Additional information}

\textbf{Competing financial interests:} The authors declare no competing financial interests.

\end{document}